\documentclass[aps,showpacs,preprint,nofootinbib,amsmath,amssymb,12pt]{revtex4}
\usepackage{bm}
\usepackage[dvips]{graphicx}
\usepackage[dvips]{color}
\usepackage{ascmac}
\usepackage{ulem}

\begin{document}

\title{\vspace*{0.5cm}
A pair of extremal charged black holes on Kerr-Taub-bolt space
}
\bigskip
\author{
${}^{1}$Ken Matsuno\footnote{E-mail: matsuno@sci.osaka-cu.ac.jp}, 
${}^{1}$Hideki Ishihara\footnote{E-mail: ishihara@sci.osaka-cu.ac.jp}, 
and 
${}^{2}$Masashi Kimura\footnote{E-mail: m.kimura@damtp.cam.ac.uk}
\bigskip
\bigskip
}
\affiliation{
${}^{1}$Department of Mathematics and Physics, Osaka City University, Sumiyoshi, 
Osaka 558-8585, Japan
\\
${}^{2}$DAMTP, University of Cambridge, Centre for Mathematical Sciences, Wilberforce Road, 
Cambridge CB3 0WA, United Kingdom
\bigskip
\bigskip
}

\begin{abstract}
We construct asymptotically Kaluza-Klein solutions in five-dimensional Einstein-Maxwell theory 
which represent a pair of extremal, charged, static black holes on Kerr-Taub-bolt space.
Regularity conditions require that the topology of spatial infinity and that of each black hole are not S$^3$,
but different lens spaces. 
We show that for a given topology at spatial infinity, 
there are an infinite number of different horizon topologies for the black hole pair.
We briefly discuss a generalization to the case with a positive cosmological constant. 
\end{abstract}

\preprint{OCU-PHYS-420}
\preprint{AP-GR-120}


\date{\today}
\maketitle

\section{Introduction}
In the context of string theories and brane world models, 
higher-dimensional spacetimes are actively discussed.  
Since the effects of extra dimensions would appear clearly 
in strong gravity regimes, 
we focus on higher-dimensional black hole spacetimes 
as a first step to reveal the higher-dimensional effects.  
If higher-dimensional black hole solutions have compactified extra 
dimensions, 
we can regard such black hole solutions as candidates of 
realistic models, 
since our observable world is effectively four-dimensional.  
We call these Kaluza-Klein black holes.

Comparing to spacetimes which consist of direct products of four-dimensional spacetimes and a compact 
extra dimension, spacetimes with a twisted compact extra dimension could admit a large isometry group.  
In such cases, we can obtain exact Kaluza-Klein black hole solutions easily.
A family of five-dimensional squashed Kaluza-Klein black hole solutions 
\cite{Dobiasch:1981vh, Gibbons:1985ac, Gauntlett:2002nw, 
Gaiotto:2005gf, Ishihara:2005dp, Ishihara:2006iv, 
Matsuno:2008fn, 
Chen:2010ih, Nedkova:2011hx, 
Tatsuoka:2011tx, 
Matsuno:2012ge, 
Tomizawa:2012nk, Stelea:2012ph, 
Kanou:2014rya}   
asymptote to effective four-dimensional spacetimes 
with a twisted S$^1$ as an extra dimension at infinity, 
and represent fully five-dimensional black holes near the squashed S$^3$ horizons  
(See \cite{Tomizawa:2011mc} as a review).

In this paper, 
we focus on Kaluza-Klein black holes in the five-dimensional spacetimes 
with a twisted compact extra dimension.  
In such spacetimes, 
even if we fix the spatial topology of infinity, and impose asymptotically locally 
flatness, 
there are various solutions of five-dimensional Kaluza-Klein black hole. 
We obtain new exact solutions that represent a pair of Kaluza-Klein black holes, 
and study such the possibility explicitly.

Because of nonlinearity of gravity, 
it is difficult to construct exact solutions 
that describe multi configuration of black holes. 
However, if we focus on the extremal charged black holes on Ricci-flat base spaces,
Einstein-Maxwell equations reduce to Laplace's equations. 
In four-dimensional Einstein-Maxwell theory, 
due to the balance of gravitational attraction and electrical repulsion,  
we can obtain exact solutions 
that represent extremal charged multi-black holes 
\cite{Majumdar, Papapetrou} by the superpositions of harmonic functions. 
Similarly, 
in five-dimensional Einstein-Maxwell theory,     
exact solutions of extremal charged multi-black holes are constructed  
in asymptotically flat spacetimes with the spatial topology of the three-dimensional sphere
\cite{Myers:1986rx, Gauntlett:1998fz}, 
in asymptotically locally flat spacetimes with the topology of the lens space 
\cite{Ishihara:2006pb, Ishihara:2006ig, Matsuno:2007ts}, 
and 
in asymptotically Kaluza-Klein spacetimes 
\cite{Gauntlett:2002nw, Myers:1986rx, Maeda:2006hd, Ishihara:2006iv, Matsuno:2008fn, 
Matsuno:2012ge}.
These five-dimensional multi-black hole solutions are constructed on four-dimensional 
Ricci-flat base spaces with different asymptotic structures.

In the five-dimensional multi-black hole spacetimes,
the topology of each black hole is related to the topology of spatial infinity. 
For example, 
in the multi-black hole solutions on flat base space \cite{Myers:1986rx},  
both the topology of each black hole and that of spatial infinity are S$^3$.   
On the other hand, 
in the multi-black hole solutions 
on Gibbons-Hawking base spaces 
\cite{Gauntlett:2002nw,Ishihara:2006iv,Matsuno:2008fn, Matsuno:2012ge,
Ishihara:2006pb, Ishihara:2006ig, Matsuno:2007ts},   
the topology of the $i$-th black hole is S$^3$
or the lens space $L(p_i ; 1)$ $(\text{natural numbers}~ p_i \neq 1)$, 
and 
the topology of spatial infinity becomes the lens space $L(\sum_i p_i ; 1)$. 
In the present paper, 
we find new exact solutions where the rich variety of black hole topologies is allowed 
even if the topology of the spatial infinity is fixed.

We construct exact solutions 
in five-dimensional Einstein-Maxwell theory that describe 
a pair of extremal, charged, static black holes 
on Kerr-Taub-bolt base space 
\cite{Gibbons:1979nf}, which is a generalization of the base spaces used in Refs. 
\cite{Gauntlett:2002nw, Ishihara:2006iv, Matsuno:2008fn, 
Chen:2010ih, Stelea:2012ph, Matsuno:2012ge, Kunz:2008rs, Kunz:2013osa, 
Ishihara:2006pb, Ishihara:2006ig, Matsuno:2007ts}.
As discussed in Refs. \cite{Ghezelbash:2007kw,Chen:2010zu},  
Kerr-Taub-bolt space has conical singularities at the poles on the bolt.  
By using harmonic functions on Kerr-Taub-bolt space, we can put black holes 
on these points so that the conical singularities are concealed behind 
the horizons.
Therefore, we can obtain five-dimensional black hole spacetimes that are regular 
on and outside the horizons.  
Inspecting the regularity, we find that 
the topology of each black hole and the topology of spatial infinity 
are not S$^3$ but lens space.

Kunduri and Lucietti found the black lens, black hole solutions with lens space topologies, 
with asymptotically globally flatness \cite{Kunduri:2014kja} 
by introducing solitonic disk objects and Chern-Simons term.  
On the other hand, we obtain regular multi-black lens solutions 
where asymptotic spatial topology is a different lens space
for pure Einstein-Maxwell system.
The multi-black hole solutions on Gibbons-Hawking space are known as  
the black lens solutions in asymptotic lens space topology \cite{
Gauntlett:2002nw, Ishihara:2006iv, Matsuno:2008fn, Matsuno:2012ge, 
Ishihara:2006pb, Ishihara:2006ig, Matsuno:2007ts}. 
These solutions admit only finite numbers of possibility for the topologies
of black lens if the topology of infinity is fixed.  
In contrast,    
the present new solutions allow infinite sequences of lens space 
topologies of black holes. 
This shows that   
there exist rich possible topologies of Kaluza-Klein black hole spacetimes 
even if the topology of spatial infinity is fixed.

This paper is organized as follows. 
We present explicit forms of solutions in section \ref{sec:solutions}.
Some known multi-black hole solutions are obtained by taking limits in our solutions.  
We show analytic extensions across the black hole horizons in section \ref{sec:extension}. 
In section \ref{subsec:regularity}, 
we discuss regularity conditions and topology of solutions with an explicit example. 
We investigate possible horizon topologies under fixed topology of infinity 
in section \ref{sec:posshortop}. 
We give summary and discussion, and 
generalize our solutions to the solutions with a positive cosmological constant 
in section \ref{sec:summary}.

\section{Solutions}
\label{sec:solutions}

We construct a pair of extremal, charged, static black holes on Kerr-Taub-bolt space as 
exact solutions in five-dimensional Einstein-Maxwell theory with the action 
\begin{align}
 	S = \frac{1}{16\pi} \int d^5 x \sqrt{-g} \left( R - F_{\mu\nu} F^{\mu\nu} \right) .
\label{eq:action}
\end{align}
The metric and the Maxwell field are given by 
\begin{align}
ds^2 &= - H (r , \theta)^{-2} dt ^2 + H (r , \theta) ds_4 ^2  ,
\label{eq:metric}
\\
A_\mu dx^\mu &= \pm \frac{\sqrt 3}{2} H (r , \theta)^{-1} dt ,
\label{eq:field}
\end{align}
with 
\begin{align}
ds^2_4 = & ~\Xi (r,\theta ) \left[ \frac{ {dr}^2}{\Delta(r)} + {d\theta}^2 \right]
+\frac{\sin ^2 \theta }{\Xi (r,\theta )} 
\left[ 
2 \alpha \nu d\psi - (r^2 -\nu^2 - \alpha ^2) d\phi
\right]^2
\notag \\
& 
+ \frac{\Delta (r)}{\Xi(r,\theta )}
\left[ 
2 \nu d\psi + (2 \nu \cos \theta + \alpha \sin ^2 \theta) d\phi
\right]^2 ,
\label{base} 
\\
\Delta (r) = & ~r^2 -2 \mu r + \nu ^2 - \alpha ^2 ,
\quad
\Xi (r, \theta ) 
= r^2 -(\nu - \alpha  \cos \theta )^2 , 
\label{basefuncs} 
\end{align} 
and 
\begin{align}
H(r , \theta ) = 
 1  + \frac{m_+}{r - \mu - (r_b - \mu) \cos \theta} 
 + \frac{m_-}{r - \mu + (r_b - \mu) \cos \theta} ,
\label{hfunction2} 
\end{align}
where 
$m_\pm,~ \alpha ,~\mu ,~\nu ,~r_b := \mu + \sqrt{\mu ^2 - \nu ^2 + \alpha ^2}$ are constants. 
Equation \eqref{base} is the metric 
of four-dimensional Euclidean Kerr-Taub-bolt space \cite{Gibbons:1979nf}, 
which is Ricci flat.\footnote{
Appearance of conical singularities in four-dimensional 
Kerr-Taub-bolt space is discussed in Refs. \cite{Ghezelbash:2007kw, Chen:2010zu}.  
}
Kerr-Taub-bolt space has
a bolt at $r = r_b$, where $\Delta (r = r_b) = 0$.\footnote{
A set of fixed points of a spatial Killing vector field, now it is 
$\partial/\partial \psi 
+ 2 \alpha \nu/(r_b^2 - \nu ^2 - \alpha ^2) \partial/\partial \phi$, 
is called a bolt if the set is two-dimensional manifold.
}
If we require that the metric and the Maxwell field take the forms of \eqref{eq:metric} and \eqref{eq:field}, 
and the four-dimensional space $ds^2 _4$ is Ricci flat,  
Einstein-Maxwell equations reduce to Laplace's equations.  
The function $H(r,\theta)$ in the form of \eqref{hfunction2} 
is a harmonic function on Kerr-Taub-bolt space \eqref{base}.\footnote{
The harmonic function \eqref{hfunction2} is a generalization of 
the harmonic functions on Taub-bolt space \cite{Nedkova:2011hx} 
and Eguchi-Hanson space \cite{Ishihara:2006pb}. 
}   
Two black holes are located
at the north pole ($\theta = 0$) and the south pole ($\theta = \pi$)
on the bolt.

The coordinates 
run the ranges of 
$- \infty < t < \infty, ~ r_b < r < \infty$, and $0 \leq \theta \leq \pi$. 
The ranges of $\phi$ and $\psi$ will be determined from the regularity conditions of the spacetime 
in section \ref{subsec:regularity}.
For the spacetime signature to be $(- , + , + , + , +)$, 
the inequalities $r_b > 0$ and $\Xi (r, \theta ) > 0$ should hold.  
Then we have the inequalities $\mu > 0$ and $\mu > |\nu|$. 
Since we can consider the case $\alpha > 0$ and $\nu > 0$ 
by rearrangement of the angular coordinates 
without loss of generality, 
in this paper, 
we restrict ourselves to 
the ranges of parameters such that\footnote{
The solution \eqref{eq:metric} is not a supersymmetric solution  
since Kerr-Taub-bolt base space \eqref{base} is not 
a hyper-K\"{a}hler space in the ranges of parameters \eqref{parahanni}. 
}
\begin{align}
m_\pm > 0 , \quad \alpha > 0 , \quad \mu > \nu > 0. 
\label{parahanni}
\end{align}

\subsection{Limits}
By taking some limits in the solution \eqref{eq:metric},  
we obtain some known, extremal, charged, static, asymptotically locally flat, regular black hole solutions.  
First, 
when $\alpha = 0 ,~ \mu = \nu$,  
the
bolt shrinks to a NUT singularity, 
and 
the solution coincides with the single black hole solution on self-dual Taub-NUT base space 
\cite{Gauntlett:2002nw, Ishihara:2005dp}. 
Secondly, 
when $\alpha = 0 ,~ \mu = \nu \left[ 1 + a^4 / (128 \nu ^4) \right]$,  
then taking the limit $\nu \to \infty$ with $\tilde r ^2 = 4 (r ^2 - \nu ^2)$ held fixed,  
the solution 
represents a pair of black holes on Eguchi-Hanson base space \cite{Ishihara:2006pb}. 
Thirdly, 
when $\alpha = \nu = 0$,   
the solution 
represents a pair of black holes 
on a Kaluza-Klein bubble (equivalently, on Euclidean Schwarzschild base space) 
\cite{Kunz:2008rs, Kunz:2013osa}. 
Fourthly, 
when $\alpha = 0$, 
the solution 
represents a pair of 
black holes on
Taub-bolt base space \cite{Stelea:2012ph}. 
Lastly, 
when $\nu =0$, 
the solution represents a pair of black holes on Euclidean Kerr base space \cite{Chen:2010ih}.

\subsection{Asymptotic behavior}
The asymptotic behavior of the metric \eqref{eq:metric} near the infinity $r \to \infty$ becomes 
\begin{align} 
 ds^2 \simeq &
-\left(1 - 2\frac{m_+ + m_-}{r}\right)dt^2 
+ 
\left(1 + \frac{m_+ + m_- +2\mu}{r}\right) dr^2 
\notag\\ & + 
r^2 \left(1 + \frac{m_+ + m_- }{r}\right) \left( d\theta^2 + \sin^2 \theta {d\phi}^2 \right) 
\notag\\ & +  4 \nu ^2
\left(1 + \frac{m_+ + m_- -2\mu}{r}\right)
 \left( d \psi + \cos \theta d\phi \right)^2.
\label{enpoumetric}
\end{align} 
We see that this solution is asymptotically locally flat, i.e., 
the metric asymptotes to a twisted constant S$^1$ fiber bundle over the four-dimensional 
Minkowski spacetime, 
and the parameter $\nu$ denotes the size of extra dimension.
However, we cannot take the limit $\nu \to \infty$ while keeping the radius of the bolt finite.
This is because $r_b = \mu + \sqrt{\mu ^2 - \nu ^2 + \alpha ^2}$ is larger than $\nu$ if $\mu > \nu >0$. 

As is seen later, 
we need to require that 
the topology of $r = {\rm const.} (>r_b)$ surface is a lens space 
$L(p; q)$ for coprime natural numbers $p$ and $q$ with $p > q > 0$ 
(see Appendix \ref{appendix:lensspace} for the definition of a lens space).
The Komar mass $M$ and the total electric charge $Q$ at the infinity are given by  
\begin{align} 
 M = \frac{\sqrt 3}{2} |Q| = 
\frac{3 \nu}{\pi } \left( m_+ + m_- \right) \frac{\mathcal A_{\rm S^3}}{p}, 
\end{align}  
where $\mathcal A_{\rm S^3} = 2\pi^2$ denotes the area of a unit S$^3$.

\section{Analytic extensions across the black hole horizons}
\label{sec:extension}
In this section, we show that
two black holes are located at the north pole ($\theta = 0$) and the south pole ($\theta = \pi$) 
on the 
bolt ($r = r_b$). 
In the coordinate system $(t, r, \theta , \phi , \psi )$, 
the metric \eqref{eq:metric} with 
\eqref{base}-\eqref{parahanni} 
diverges apparently 
at $r = r_b , ~ \theta = 0$ and $r = r_b , ~ \theta = \pi$.

\begin{figure}[htbp]
\centering
\includegraphics[width=7cm]{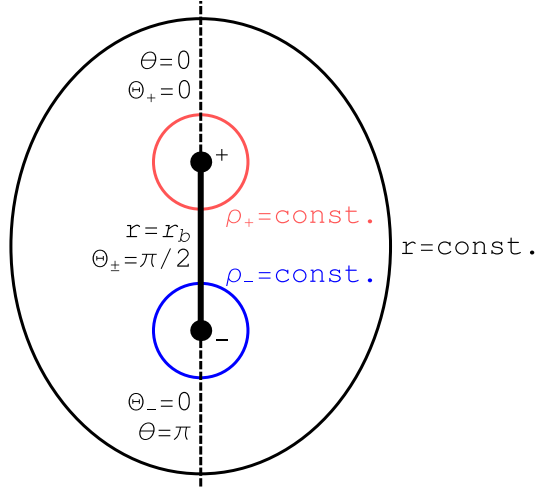}
\caption{
The relation between the coordinates $(r , \theta)$ and $(\rho_\pm , \Theta_\pm)$. 
The bolt $r = r_b$ is shown by the thick segment. The northern black hole on the 
bolt is denoted by the dot \lq$+$\rq\, and the southern one by the dot \lq$-$\rq . 
The symmetric polar axes $\theta = 0$ and $\theta = \pi$ are denoted by 
the broken lines, respectively. 
\bigskip}
\label{fig:zahyouhenkan}
\end{figure}

In order to remove these coordinate singularities
we introduce new coordinates 
$(v_\pm , \rho_\pm , \Theta_\pm , \Phi_\pm , \Psi_\pm)$ such that
\begin{align}
 & dv_\pm - F_\pm (\rho_\pm, \Theta_\pm) d\rho_\pm
\notag \\
& \qquad
- d\Theta_\pm
\left(
\sqrt{ \frac{m_\pm ^3 (r_b \mu + \nu (-\nu \pm \alpha) ) }{(r_b - \mu )^3} }
\sin \Theta _\pm \cos ^3 \Theta _\pm
\right.
\cr
&\hspace{3cm} \left. + \int d\rho_\pm \partial_{\Theta_\pm} F_\pm (\rho_\pm, \Theta_\pm)
\right)
= dt,
\\
 &4 (r_b - \mu ) \rho_\pm \cos^2 \Theta_\pm = r -  r_b,
\\ &
8 \rho_\pm \sin^2 \Theta_\pm = \left(\theta - \frac{\pi}{2} \pm
\frac{\pi}{2} \right)^2,
\\
&\Phi_\pm =
\pm \frac{r_b ^2 - \nu ^2 - \alpha ^2}{r_b ^2 - (\nu \mp \alpha) ^2}
\left(\frac{2 \alpha \nu}{r_b ^2 - \nu ^2 - \alpha ^2 } \psi - \phi \right) ,
\label{phipm}
\\
 &\Psi_\pm = \frac{2 \nu (r_b -\mu)}{r_b ^2 - (\nu \mp \alpha) ^2}
(\psi \pm \phi ) ,
\label{psipm}
\end{align}
where the functions $F_\pm(\rho_\pm, \Theta_\pm)$ are defined by
\begin{align}
 F^2_\pm (\rho_\pm, \Theta_\pm) 
= \frac{2 H^3 \Xi }{\Delta  \rho _\pm}
\left( \Delta  \sin ^2 \Theta _\pm +8 \rho _\pm (r_b-\mu )^2 \cos ^4
\Theta _\pm \right) ,
\end{align}
and $0 \leq \Theta_\pm \leq \pi/2$.
We show the relation between the coordinates $(r, \theta)$ and $(\rho_\pm , \Theta_\pm)$ 
in Fig.\ref{fig:zahyouhenkan}.

In these coordinates, the metric \eqref{eq:metric} behaves as 
\begin{align}
 ds^2 \simeq & 
 8 \sqrt{ \frac{ (r_b - \mu)(r_b \mu + \nu(- \nu \pm \alpha ))}{m_\pm} } 
dv_\pm d\rho_\pm
 \notag \\
 & 
 + \frac{4 m_\pm (r_b \mu + \nu(- \nu \pm \alpha )) }{r_b - \mu}
 \left( d\Theta_\pm ^2 + \sin^2 \Theta_\pm d\Phi_\pm ^2 + \cos^2 \Theta_\pm d\Psi_\pm ^2 \right) 
+ O (\rho_\pm) ,
\label{extension}
\end{align}
near $\rho_\pm = 0$. The metric well behaves at the null surfaces $\rho_\pm = 0$. 
We see that $\eta = \partial/\partial t = \partial / \partial v_\pm$ 
is a Killing vector field that becomes null at $\rho_\pm = 0$,  
and $\eta$ is hypersurface orthogonal to the surfaces $\rho_\pm = 0$.
These mean that the null hypersurfaces $\rho_\pm = 0$ are Killing horizons. 
From the next-order terms of the metric, we see the norm of the Killing vector field 
$\partial / \partial v_\pm$ is proportional to $\rho_\pm ^2$ near $\rho_\pm = 0$. 
Then the surface gravity vanishes, and the solution is extremal.   
Hence the solution \eqref{eq:metric} with   
\eqref{base}-\eqref{parahanni}  
describes  
a pair of extremal charged black holes on Kerr-Taub-bolt base space.

In fact, one can easily check that 
the set of new coordinates $(v_\pm , \rho_\pm , \Theta_\pm , \Phi_\pm , \Psi_\pm)$ 
gives analytic extensions across the horizons.
So the spacetime has no curvature singularity on and outside the horizons.

\section{regularity conditions and topology}
\label{subsec:regularity}
In this section we discuss the regularity conditions for the absence of conical singularities
on the spacetime. 
The metrics of angular parts of the black hole horizons 
in equation (\ref{extension})
are locally round ${\rm S}^3$.  
In order to remove the conical singularities at $\Theta_\pm=0, \pi/2$, 
we should impose the periodic conditions 
\begin{align}
 \left(\Phi_\pm, \Psi_\pm \right)\sim\left(\Phi_\pm + 2 \pi, \Psi_\pm \right)
\label{idoriginal1}
\end{align}
at $\Theta_\pm=0$, and
\begin{align}
\left(\Phi_\pm, \Psi_\pm \right)\sim\left(\Phi_\pm, \Psi_\pm + 2 \pi \right)
\label{idoriginal2}
\end{align}
at $\Theta_\pm=\pi/2$.

We translate the conditions for $\left(\Phi_\pm, \Psi_\pm \right)$ to a coordinate system 
which covers the regions around the north pole and the south pole simultaneously.
Via the original coordinate system $(\phi, \psi)$,
we introduce such the coordinates as
\begin{align}
\Phi =\frac12(\psi-\phi),\quad \Psi=\frac12(\psi+\phi).
\label{global_coord}
\end{align}
Using the relations \eqref{phipm}, \eqref{psipm}, and \eqref{global_coord} we have 
\begin{align}
\Phi_+=\Phi- \frac{\sigma_1}{\sigma_2} \Psi, \quad 
&\Psi_+= \frac{1}{\sigma_2} \Psi, 
\label{phi_psi_+}
\\
\Phi_-=  \Psi- \frac{\sigma_2}{\sigma_1}\Phi, \quad
&\Psi_-= \frac{1}{\sigma_1} \Phi, 
\label{phi_psi_-}
\end{align}
where $\sigma_1$ and $\sigma_2$ are defined by
\begin{align}
\sigma_1=\frac{r_b^2-(\nu+\alpha)^2}{4\nu(r_b-\mu)}, \quad  
\sigma_2=\frac{r_b^2-(\nu-\alpha)^2}{4\nu(r_b-\mu)}. 
\label{id_point}
\end{align}
We note that $\sigma_1 >0$ and $\sigma_2 >0$ in the ranges of parameters \eqref{parahanni}.
The four conditions \eqref{idoriginal1} and \eqref{idoriginal2} reduce to 
the three conditions 
\begin{align}
& \left( \Phi, \Psi \right) \sim \left( \Phi, \Psi + 2\pi \right),
\label{NP_reg1}
\\
& \left( \Phi, \Psi \right) \sim \left( \Phi+ 2\pi, \Psi  \right),
\label{NP_reg2}
\end{align}
and 
\begin{align}
\left( \Phi, \Psi \right) \sim \left( \Phi+2\pi \sigma_1, \Psi + 2\pi \sigma_2 \right).
\label{SP_reg}
\end{align}
The regularity conditions \eqref{idoriginal1} at the north pole axis of the northern black hole 
($\Theta_+=0$), 
and at the north pole axis of the southern black hole ($\Theta_-=0$) correspond to 
the regularity conditions \eqref{NP_reg1} and \eqref{NP_reg2} at the north pole axis 
($\theta=0$) 
and the south pole axis ($\theta=\pi$) of the $r=\mbox{const.}(>r_b)$ surface. 
On the other hand, the regularity conditions \eqref{idoriginal2} at the south pole axes  
of two black holes ($\Theta_\pm=\pi/2$) reduce to one condition \eqref{SP_reg} which corresponds to 
the regularity condition on the bolt off the horizons 
(see Fig.\ref{fig:zahyouhenkan} and Appendix \ref{appendix:boltregularity}).

From the inequalities \eqref{parahanni} for $\alpha,~ \mu$, and $\nu$ 
we see that 
\begin{align}
0<\sigma_2-\sigma_1<1< \sigma_1+\sigma_2 .
\label{s_region}
\end{align}
If $\sigma_1$ and $\sigma_2$ were $0$ or integers which are coprime, 
the metric \eqref{eq:metric} with the regularity 
conditions \eqref{NP_reg1} and \eqref{NP_reg2} would mean that the $r = {\rm const.} ( >  r_b)$ 
surface is topologically S$^3$. 
However, the inequality \eqref{s_region} does not permit the case. 
The condition \eqref{SP_reg} with \eqref{s_region} requires that 
the point $(\Phi, \Psi)=(0, 0)$ should be identified with a point in the region 
$\Phi<\Psi<\Phi+2\pi$, and $2\pi<\Phi+\Psi$. 
Namely, the condition \eqref{SP_reg} is an additional identification,   
then we should consider the case that the $r= {\rm const.} ( > r_b)$ surface is a lens 
space $L(p;q)$, 
which is specified by the additional identification condition 
\begin{align}
	\left(\Phi, \Psi\right) 
	\sim \left( \Phi+ 2 \pi \frac{q}{p}, \Psi + 2 \pi \frac{1}{p} \right),
\label{r_const_lens}
\end{align}
where $p, q$ are natural numbers which are coprime.
The identification \eqref{r_const_lens} in the present paper is graphically shown 
in Appendix \ref{appendix:lensspace}. 
We should note that the lens spaces $L(p;q)$ and $L(p;p-q)$ are topologically equivalent, 
however, these spaces with the metric \eqref{eq:metric} are different in geometry.

We show all the conditions \eqref{NP_reg1}, \eqref{NP_reg2}, \eqref{SP_reg}, and 
\eqref{r_const_lens} are compatible if the parameters $\alpha$ and $\mu$ are 
chosen suitably. 
In order that the conditions \eqref{SP_reg} and \eqref{r_const_lens} are compatible, 
$\sigma_1$ and $\sigma_2$ are given by 
\begin{align}
	\sigma_1= \frac{k_1}{p}, \quad \sigma_2= \frac{k_2}{p},
\end{align}
where $k_1,~ k_2$, and $p$ are natural numbers which are coprime to each other. 
If these are not coprime, there appears a shorter period of identification 
so that a conical singularity emerges. 
The inequality \eqref{s_region} means
\begin{align}
0<k_2-k_1<p < k_1+k_2.
\label{param_region}
\end{align}
Since $k_2$ and $p$ are coprime, there exists a natural number $n$ 
such that $(n k_2 \bmod p) = 1$.\footnote{
For integers $a$ and $b (>0)$, we understand that 
the remainder of the division of $a$ by $b$ is denoted by $(a \bmod b)$, 
where $0\leq (a \bmod b) \leq b-1$.}
By the use of the number $n$, the parameter $q$ is given by 
\begin{align}
q = (n k_1 \bmod p).
\label{lens_q}
\end{align}
It is easy to see that $p$ and $q$ are coprime.

Rewriting the conditions \eqref{NP_reg1}, \eqref{NP_reg2}, and \eqref{SP_reg}  
in the coordinates 
$\left(\Phi_\pm, \Psi_\pm \right)$  we have the conditions \eqref{idoriginal1}, 
\eqref{idoriginal2}, and  
\begin{align}
& \left(\Phi_+ , \Psi_+ \right) 
\sim \left( \Phi_+ - 2 \pi \frac{k_1}{k_2},~ \Psi_+ + 2 \pi\frac{p}{k_2} \right),
\label{NP_lens}\\
& \left(\Phi_- , \Psi_- \right) 
\sim \left( \Phi_- - 2 \pi \frac{k_2}{k_1},~ \Psi_- + 2 \pi\frac{p}{k_1} \right). 
\label{SP_lens}
\end{align}
If the conditions \eqref{NP_lens} and \eqref{SP_lens} were absent, 
the topologies of black holes would be S$^3$, but in fact the existence of them 
in addition to \eqref{idoriginal1} and \eqref{idoriginal2} requires that 
the topologies of the northern and the southern black hole horizons are $L(p_+; q_+)$ and 
$L(p_-; q_-)$, 
 where $(p_+ , q_+)$ and $(p_- , q_-)$ denote two sets of coprime natural numbers.
Here, two points $(\Phi, \Psi)=(q/p, 1/p)$ and  $(1/p, \tilde q/p)$ 
directly determine the topological parameters, 
where $\tilde q$ is the natural number less than $p$ such that 
\begin{align}
(q \tilde q \bmod p)=1.
\label{tilde_q}
\end{align}  
Substituting $(\Phi, \Psi)=(q/p, 1/p)$ into 
\eqref{phi_psi_+}, and  $(\Phi, \Psi)=(1/p, \tilde q/p)$ into 
\eqref{phi_psi_-}, we have
\begin{align}
(\Phi_+, \Psi_+)=\left(\frac{1}{k_2}\frac{1}{p}(qk_2-k_1), \frac{1}{k_2}\right)
=\left(\frac{q_+}{p_+}, \frac{1}{~p_+}\right),
\\
(\Phi_-, \Psi_-)=\left(\frac{1}{k_1}\frac{1}{p}(\tilde qk_1-k_2), \frac{1}{k_1}\right)
=\left(\frac{q_-}{p_-}, \frac{1}{~p_-}\right).
\end{align}
Then the lens parameters $p_\pm$ and $q_\pm$ are given by 
\begin{align}
&p_+=k_2,\quad 
q_+=\frac{1}{p}(q k_2 - k_1), 
\label{pq_+}
\\
& p_-=k_1,\quad  
q_-=\frac{1}{p}(\tilde q k_1 - k_2).
\label{pq_-}
\end{align}
From \eqref{param_region}, \eqref{pq_+}, and \eqref{pq_-}, we see that the lens parameters 
of the northern and the southern black holes and spatial infinity, $p_\pm$ and $p$, are 
coprime to each other, and satisfy
\begin{align}
0< p_+-p_-<p< p_++p_- . 
\end{align}

Solving \eqref{id_point} for $\alpha$ and $\mu$, we have 
\begin{align}
	\frac{\alpha}{\nu} &=
	\frac{2 p}{k_2 - k_1}
\sqrt \frac{k_1 k_2}{p^2 - (k_2 - k_1)^2} - \frac{k_1 + k_2}{k_2 - k_1} ,
\label{alphawarunu}
\\
	\frac{\mu}{\nu} &=
	\frac{p(k_1 + k_2)-2 \sqrt{k_1 k_2 \left( p^2-(k_2-k_1)^2 \right)}}{(k_2 - k_1)^2}.
\label{muwarunu}
\end{align}
The physical parameters $\alpha/\nu$ and 
$\mu/\nu$ take discrete values. 
Then $r_b/\nu$ also has a discrete value
\begin{align}
\frac{r_b}{\nu} &= 2 \sqrt{ \frac{k_1 k_2}{p^2 -(k_2-k_1)^2} } .
\label{r_b}
\end{align}
The set of three natural numbers $p$, $k_1$, and $k_2$ determines 
the lens parameters $q ,~ p_+ ,~ q_+ ,~ p_-$, and $q_-$. 
That is, the topology of the $r = {\rm const.}$ surface, $L(p; q)$, the northern black hole 
horizon, $L(p_+; q_+)$, and  the southern black hole 
horizon, $L(p_-; q_-)$, are specified by $p$, $k_1$, and $k_2$.  
The areas of the northern and the southern black holes with lens space topologies are given by 
\begin{align}
\mathcal A_\pm 
= & \left[
 \frac{4 m_\pm (r_b \mu + \nu(- \nu \pm \alpha )) }{r_b - \mu}
 \right]^{3/2} 
\frac{1}{p_\pm} \mathcal A_{\rm S ^3},
\end{align}
respectively.

Since $p_+ = k_2 >1 $ and $p_- = k_1 > 1$, the topologies of black holes cannot be S$^3$
if $\alpha > 0,~  \mu > \nu > 0$.\footnote{
If we take the limit $\alpha \to 0$, the base space reduces to Euclidean Taub-bolt space which 
is regular everywhere. In this case, the topology of black hole can be S$^3$.}
This implies that conical singularities appear if we take no black hole limit $m_\pm \to 0$.
This result is consistent with the discussions in Refs. \cite{Ghezelbash:2007kw,Chen:2010zu}.
Even though Euclidean Kerr-Taub-bolt space \eqref{base}
has conical singularities at the north and the south poles on the bolt,
we can obtain regular multi-black hole solution \eqref{eq:metric}  
with \eqref{base}-\eqref{parahanni} by putting black holes on them.\footnote{
This method is similar to the constructions of the multi-black hole solutions 
on Gibbons-Hawking space~\cite{Gauntlett:2002nw,Ishihara:2006iv,Matsuno:2008fn, 
Ishihara:2006pb, Ishihara:2006ig, Matsuno:2007ts}.}

\subsection{Example: The case of given $p , k_1 , k_2$}
\label{example1}
Let us consider the case of 
$p=3, k_1=4, k_2=5$ as an example.
We determine the lens parameters $p, q$, and $p_\pm, q_\pm$ by 
\eqref{lens_q}, \eqref{pq_+}, and \eqref{pq_-}. 

First, we see the lens parameter for 
the $r=\mbox{const.}(>r_b)$ surface. 
We find $n=2$ satisfies 
\begin{align}
& (n k_2 \bmod p) = (n\times 5 \bmod 3) = 1. 
\end{align}
Then, we obtain
\begin{align}
& q = (n k_1 \bmod p) = (2\times 4 \bmod 3) = 2. 
\end{align}
The topology of the $r=\text{const.}(>r_b)$ surface is $L(3; 2)$.

Next, we see the black hole horizons. 
We see that 
\begin{align}
& p_+=k_2=5, 
\quad
q_+ =\frac{1}{p}(q k_2 - k_1)=\frac{1}{3}(2\times 5 - 4)=2. 
\end{align}
The topology of the northern black hole horizon is $L(5; 2)$.

Similarly, 
we find $\tilde q=2$ solves
\begin{align}
&(q \tilde q \bmod p) = (2 \tilde q \bmod 3) = 1. 
\end{align}
Then, we have
 \begin{align}
& p_-=k_1=4,
\quad
q_- =\frac{1}{p}(\tilde q k_1 - k_2)=\frac{1}{3}(2\times 4 - 5)=1. 
\end{align}
The topology of the southern black hole horizon is $L(4; 1)$.

\begin{figure}[!htbp]
\centering
\includegraphics[width=7cm]{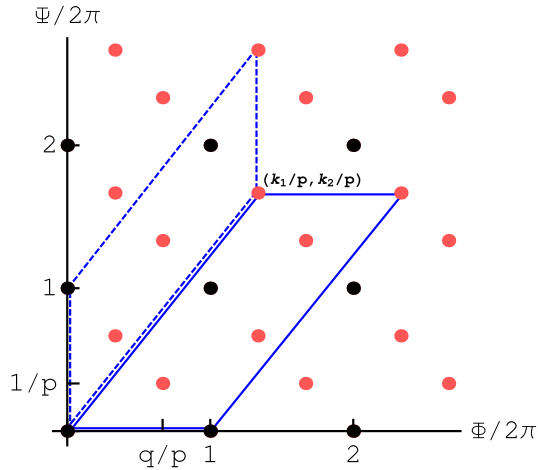}
\caption{
The identifications in the case of $p=3, k_1=4, k_2=5$.  
All points which are identified with $(0,0)$ are plotted in the $\Phi/2\pi$-$\Psi/2\pi$ plane. 
The points described by \eqref{NP_reg1} and \eqref{NP_reg2} are shown by black dots, 
and the points for \eqref{SP_reg} are by dark (red) dots.  
We see that $q=2$ in this case, then the topology of the $r = \mbox{const.}(>r_b)$ surface 
is $L(3; 2)$. 
The solid and dashed parallelograms correspond to the squares  
$\big((0,0), (1,0), (1,1), (0,1)\big)$ 
in the $\Phi_+/2\pi$-$\Psi_+/2\pi$ and $\Phi_-/2\pi$-$\Psi_-/2\pi$ planes, respectively 
(see Fig.\ref{fig:52}).  
}
\label{fig:3252}
\end{figure}

\begin{figure}[!ht]
\centering
\includegraphics[width=7cm]{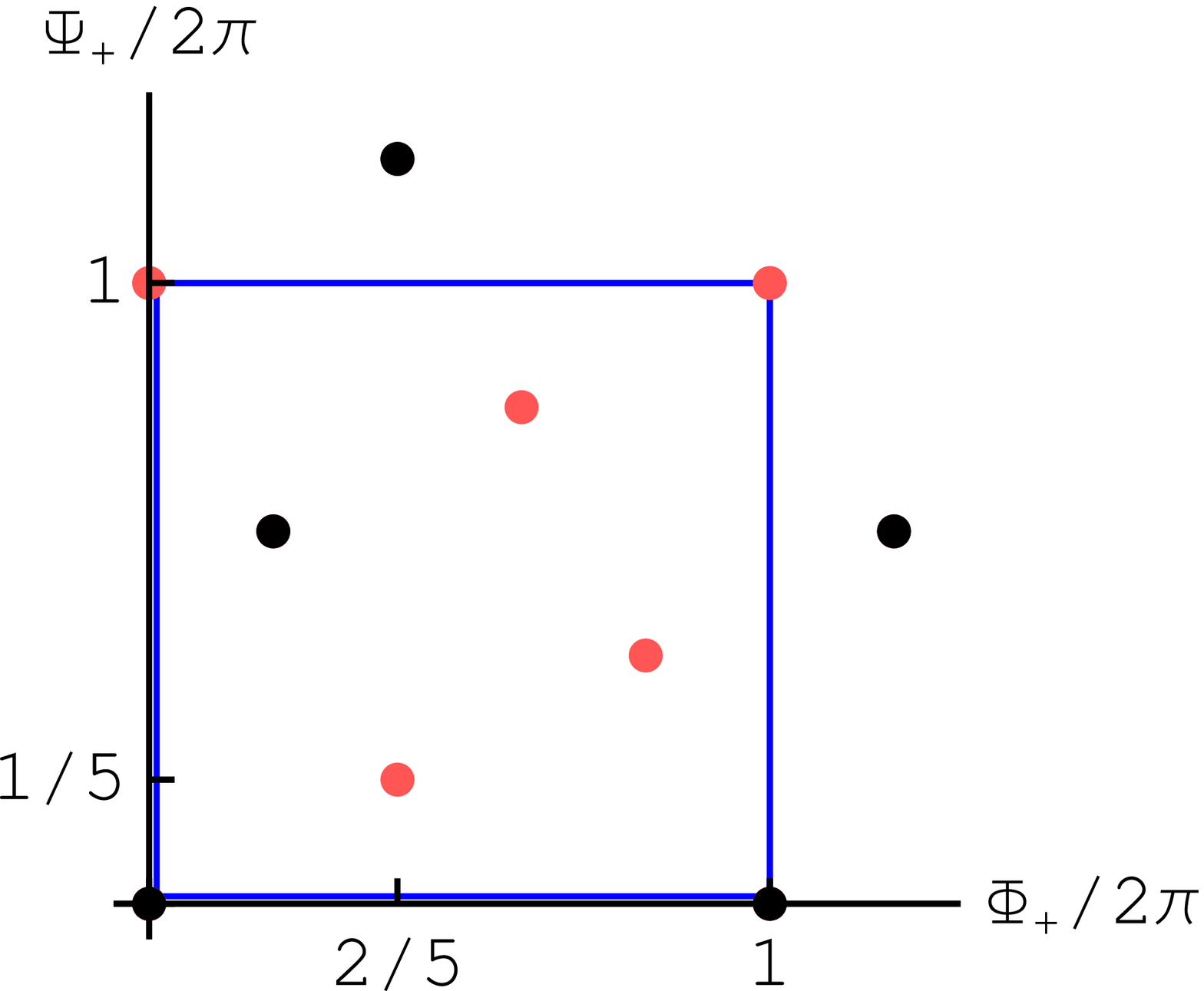}
\includegraphics[width=7cm]{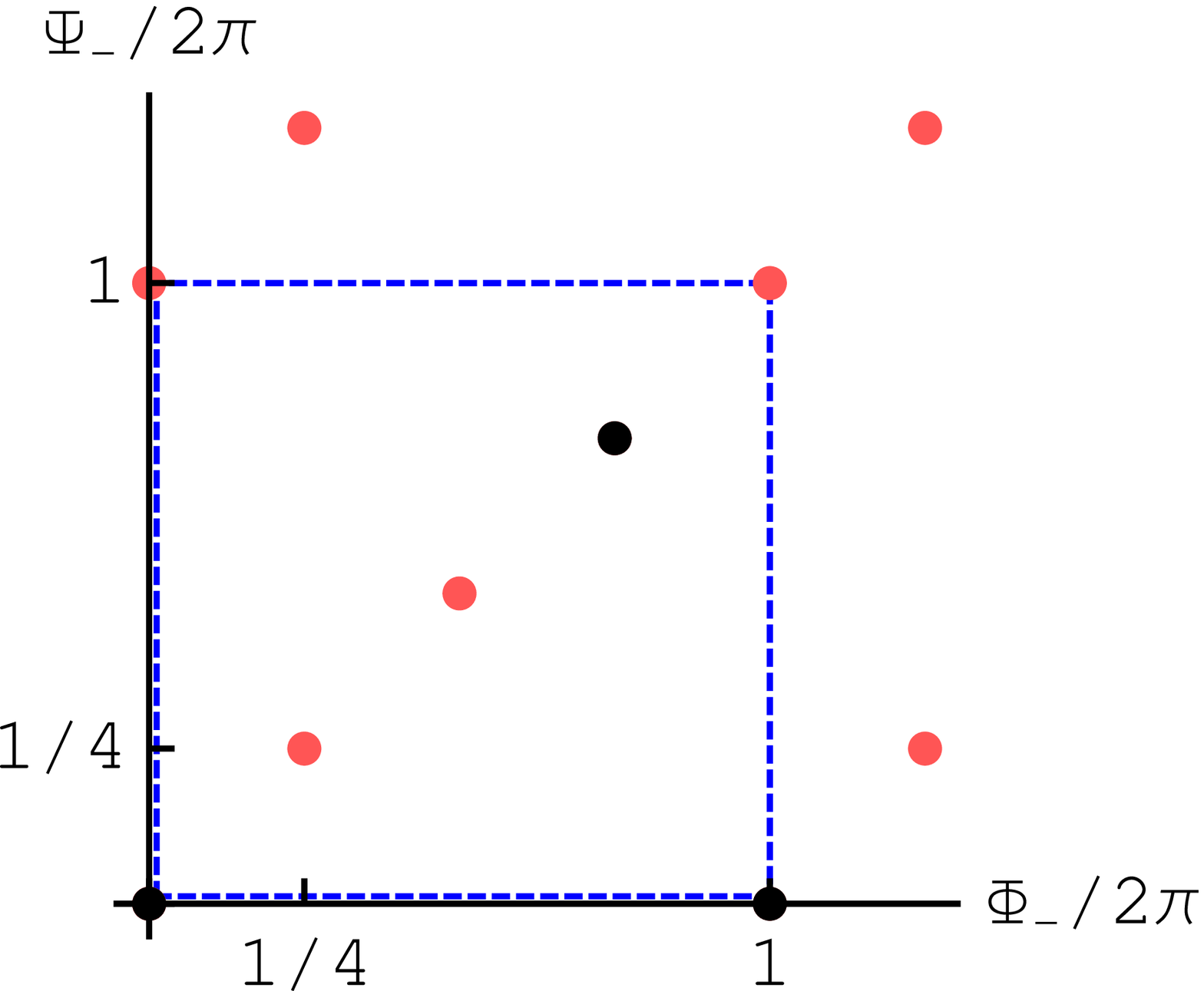}
\caption{ 
The points which are identified with $(0,0)$ are plotted in the $\Phi_+/2\pi$-$\Psi_+/2\pi$ 
plane (the left panel). The solid square corresponds to the solid parallelogram 
in Fig.\ref{fig:3252}. 
We see that the topology of the northern black hole horizon is $L(5; 2)$.  
The same is shown in the $\Phi_-/2\pi$-$\Psi_-/2\pi$ plane (the right panel). 
The dashed square corresponds to the dashed parallelogram in Fig.\ref{fig:3252}. 
The topology of the southern black hole horizon is $L(4; 1)$.   
\bigskip
}
\label{fig:52}
\end{figure}

The same thing can be understood graphically. 
Three conditions \eqref{NP_reg1}, \eqref{NP_reg2}, and \eqref{SP_reg} for the case 
$p=3, k_1=4$, and $k_2=5$ are 
compatible if the $r=\mbox{const.}(>r_b)$ surface is $L(3; 2)$ 
(see Fig.\ref{fig:3252} and Appendix \ref{appendix:lensspace}). 
Transforming identification points in the $\Phi$-$\Psi$ plane into the $\Phi_\pm$-$\Psi_\pm$ 
plane, we find the northern and the southern black hole horizons are $L(5; 2)$ and $L(4; 1)$ 
(see  Fig.\ref{fig:52}).


The parameters $\alpha, \mu$, and $r_b$ are obtained as
\begin{align}
\frac{\alpha}{\nu} 
&={6}\sqrt \frac{5}{2} - 9 ,
\quad
\frac{\mu}{\nu} 
={27- 8\sqrt{10}},
\quad
\mbox{and}\quad
\frac{r_b}{\nu}=\sqrt{10}.
\end{align}

\section{Possible horizon topology under fixed topology of infinity}
\label{sec:posshortop}
It is interesting that there are rich possibilities of lens space topologies
of the northern and the southern black hole horizons even if the topology of spatial boundary at 
infinity, equivalently, the topology of $r = {\rm const.}(>r_b)$ surface, is fixed. 
Namely, for given $p$, there are a variety of sets, $k_1$ and $k_2$, which give the same $q$. 

If we fix the topology of spatial infinity as $L(p; q)$, 
from \eqref{appendix:kl} we see that such the $k_1$ and $k_2$ are given by
\begin{align}
	k_1=a q -b p, \quad k_2= a , 
\label{k_a_b}
\end{align}
where $a$ and $b$ are natural numbers such that $a, b$, and $p$ are coprime, 
so that $k_1, k_2$, and $p$ are coprime to each other.  
Similarly, paying attention to the point $(\Phi, \Psi)=(1/p, \tilde q/p)$, we see that
\begin{align}
k_1=\tilde a, \quad k_2= \tilde a \tilde q -\tilde b p , 
\label{k_tilde_a_b}
\end{align}
where $\tilde a$ and $\tilde b$ are natural numbers such that
\begin{align}
\tilde a=a q -b p, \quad a=\tilde a \tilde q -\tilde b p,
\label{aqbp}
\end{align}
and the definition of $\tilde q$ is given by \eqref{tilde_q}.

Since the parameters $k_1$ and $k_2$ satisfy the inequality \eqref{param_region} 
then from \eqref{k_a_b} and \eqref{k_tilde_a_b} we have
\begin{align}
0< -a (q-1) +b p = \tilde a (\tilde q-1) -\tilde b p < p.  
\label{jseigen1}
\end{align}
Then, we can set 
\begin{align}
-a (q-1) +b p = \tilde a (\tilde q-1) -\tilde b p =l ,  
\label{def_l}
\end{align}
where $l$ is a natural number in the range $0<l <p$. 
Then, $a, b$ and $\tilde a, \tilde b$ satisfying \eqref{def_l} are labeled by $l$ 
as $a_l, b_l$ and $\tilde a_l, \tilde b_l$, respectively. 
Equation \eqref{def_l} means that $-b_l$ is the quotient of $-a_l (q-1)$  
devided by $p$, and $l$ is the remainder. 
At the same time, $\tilde b_l$ is the quotient of $\tilde a_l (\tilde q-1)$ 
devided by $p$, and $l$ is the remainder. 
Then, $a_l$ and $\tilde a_l$ are expressed by 
\begin{align}
(-a_l (q-1) \bmod p)=l, \quad 
(\tilde a_l (\tilde q-1) \bmod p)=l.
\label{a_l}
\end{align}
For a given value of $l$, the parameter $a_l$ and $\tilde a_l$ are infinite sequences 
of natural numbers. 
The parameters $k_1$ and $k_2$ written as \eqref{k_a_b} and \eqref{k_tilde_a_b} 
are also written by $a_l$, $\tilde a_l$, and $l$ as 
\begin{align}
k_1=\tilde a_l=a_l-l, \quad 
k_2=a_l=\tilde a_l+l. 
\label{k_12}
\end{align}
Here, $a_l$ and $\tilde a_l$ are restricted such that $k_1$ and $k_2$ are coprime. 
Furthermore, the inequality \eqref{param_region} requires 
\begin{align}
	p < 2a_l-l=2\tilde a_l+l . 
\label{jseigen2}
\end{align}

We see that there is no $a_l$ or $\tilde a_l$ for $q=1$, 
in other words, we find no $(k_1, k_2)$ in the region \eqref{param_region}. 
Otherwise, the infinite sequences of $a_l$, $\tilde a_l$ give infinite 
sequences of $k_1$, $k_2$, therefore 
there exist infinite sequences of $p_\pm$ and $q_\pm$. 
This means that the black holes can take infinite sequences of lens space topologies
even if the topology of the spatial boundary at infinity is fixed.

Substituting \eqref{def_l} and \eqref{k_12} for $a_l$, $b_l$ and $\tilde a_l$, $\tilde b_l$  
into \eqref{pq_+}, \eqref{pq_-}, we obtain\footnote{
Using $p, q, \tilde q, l, a_l$, we can write $p_\pm$ and $q_\pm$ as
$p_+=a_l,~
q_+=(l+a_l(q-1))/p,~
p_-= a_l - l,~
q_-=((a_l-l)(\tilde q-1)-l)/p$.
} 
\begin{align}
&p_+ =a_l, \quad
q_+ = b_l,
\cr
&p_- = \tilde a_l , \quad
q_- = \tilde b_l. 
\label{topopara}
\end{align}

Substituting \eqref{k_12} into \eqref{alphawarunu}, \eqref{muwarunu}, and \eqref{r_b}, 
we see the geometrical parameters are written as 
\begin{align}
\frac{\alpha}{\nu} &= 
\frac{2 p}{l} \sqrt{\frac{a_l(a_l - l)}{p^2-l^2}} - \frac{2 a_l-l}{l},
\\
\frac{\mu}{\nu} &=
\frac{1}{l^2 }\left( p( 2a_l-l) - 2 \sqrt{a_l(a_l-l) (p^2-l^2)}\right), 
\\
\frac{r_b}{\nu} &= 2\sqrt{\frac{a_l(a_l - l)}{p^2-l^2}}.
\end{align}
We see that the size of the bolt $r_b$ becomes large as $a_l$ increases. 
In this spacetime, the topologies of black holes are tightly related 
to the geometrical parameters.

\subsection{Example: The case of fixed $p, q$}
\subsubsection{ $p=4 ,~ q=3$ case}
\begin{figure}[!htbp]
\centering
\includegraphics[width=7cm]{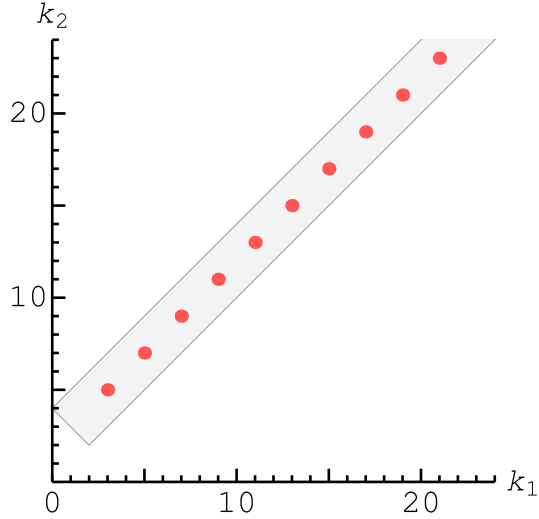}
\caption{
In the case of $p=4, q=3$, the infinite sequence of $(k_1, k_2)=(\tilde a_2, a_2)$ 
is shown by dots in the $k_1$-$k_2$ plane. 
The shaded region represents the area that satisfies the inequality \eqref{param_region}.
}
\label{fig:k1k2p4q3}
\end{figure}

In this case, we obtain $\tilde q=3$. 
From \eqref{a_l} and \eqref{k_12}, we obtain $a_l$ and $\tilde a_l$ for only $l=2$ as
\begin{align}
a_2=2j+1=k_2 , \quad \tilde a_2=2j-1=k_1, 
\quad (j=2,3,\cdots), 
\end{align}
where $j$ is restricted to satisfy the inequalities \eqref{jseigen1} and \eqref{jseigen2} with \eqref{def_l}  
(see Fig.\ref{fig:k1k2p4q3}). 
From \eqref{aqbp}, we have 
\begin{align}
b_2=j+1 , \quad \tilde b_2=j-1.
\end{align}
Then, 
\begin{align}
	p_+ = 2 j + 1 , \quad q_+ = j+1; 
\quad
	p_- = 2 j - 1 , \quad q_- = j -1 , \quad (j =2,3,\cdots).
\end{align}

The geometrical parameters $(\alpha , \mu , r_b)$ are respectively given by 
\begin{align}
\frac{\alpha}{\nu} &= \frac{2 \sqrt{4 j^2 -1}}{\sqrt{3}} -2j ,
\\
\frac{\mu}{\nu} &=4 j -\sqrt{3 (4 j^2 -1)} , 
\\
\frac{r_b}{\nu} &=\frac{\sqrt{4 j^2 -1}}{\sqrt{3}} .
\end{align}

\subsubsection{ $p=5 ,~ q=2$ case}
\begin{figure}[!htbp]
\centering
\includegraphics[width=7cm]{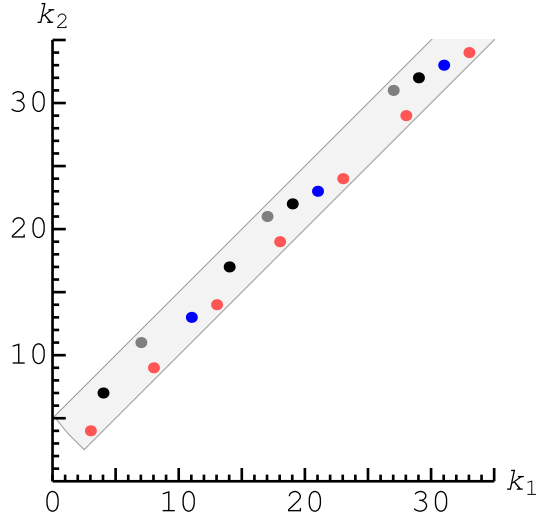}
\caption{
In the case of $p=5, q=2$, the infinite sequences of $(k_1, k_2)=(\tilde a_l, a_l) ,~ (l=1,2,3,4)$ are 
shown by dots on the line $k_2-k_1=l$ in the $k_1$-$k_2$ plane, respectively. 
}
\label{fig:k1k2p5q2}
\end{figure}

In this case, we find $\tilde q=3$ that satisfies \eqref{tilde_q}.  
There are four possible sequences of $a_l=k_2,~ \tilde a_l=k_1, (l=1,2,3,4)$ 
(see Fig.\ref{fig:k1k2p5q2}) 
satisfying \eqref{a_l} 
as
\begin{align}
	&l=1:\quad  a_1=5j-1, \quad \tilde a_1=5j-2,
\quad (j=1,2,3,\cdots), 
\cr
	&l=2:\quad a_2=5j+3, \quad \tilde a_2=5j+1,
\quad (j=2,4,6,\cdots), 
\cr 
	&l=3:\quad a_3=5j+2, \quad \tilde a_3=5j-1, 
\quad (j=1,3,4,6,\cdots, (j \bmod 3) \neq 2), 
\cr 
	&l=4:\quad a_4=5j+6 , \quad \tilde a_4=5j+2 ,
\quad (j=1,3,5,\cdots),
\end{align}
where $j$ is restricted so that $a_l$ and $\tilde a_l$ are coprime, 
and the inequalities \eqref{jseigen1} and \eqref{jseigen2} with \eqref{def_l} are satisfied.
From \eqref{aqbp}, we have
\begin{align}
	&l=1:\quad  b_1=j, \quad \tilde b_1=2j-1,
\cr
	&l=2:\quad b_2=j+1, \quad \tilde b_2=2j,
\cr 
	&l=3:\quad b_3=j+1, \quad \tilde  b_3=2j-1, 
\cr 
	&l=4:\quad b_4=j+2, \quad \tilde b_4=2j.
\end{align}
Then, from \eqref{topopara}, 
we determine $p_\pm, q_\pm$ as follows
\begin{align}
l=1:\quad 
	&p_+=5j-1, \quad q_+=j;
\qquad
	p_-=5j-2, \quad	q_-=2j-1, 
\cr 
&\quad (j=1,2,3,\cdots),
\\
l=2:\quad 
	&p_+=5j+3, \quad q_+=j+1;
\qquad
	p_-=5j+1, \quad q_-=2j, 
\cr
&\quad (j=2,4,6\cdots),
\\
l=3:\quad 
	&p_+=5j+2, \quad q_+=j+1; 
\qquad 
	p_-=5j-1, \quad q_-=2j-1, 
\cr
 &\quad (j=1,3,4,6,\cdots, (j \bmod 3) \neq 2), 
\\
l=4:\quad 
	&p_+=5j+6, \quad q_+=j+2;
\qquad 
	p_-=5j+2, \quad q_-=2j, 
\cr 
&\quad 	(j=1,3,5,\cdots).
\end{align}

\section{Summary and Discussion}
\label{sec:summary}
We construct a pair of extremal, charged, static 
black holes 
on 
Euclidean Kerr-Taub-bolt base space 
in five-dimensional Einstein-Maxwell theory. 
The metric asymptotes to an effectively four-dimensional spacetime near the infinity, 
and 
behaves as fully five-dimensional black holes near the horizons.  
Two black holes are located at the poles on the bolt.   
Each black hole has   
an analytic Killing horizon,  
then there is no curvature singularity on and outside the black hole horizons.
We note that although multi-black hole solutions in higher dimensions tend to have 
non-smooth event horizons~\cite{
Gibbons:1994vm,Welch:1995dh,Candlish:2007fh,Candlish:2009vy,
Gowdigere:2014aca,Gowdigere:2014cqa,Kimura:2014uaa}, 
our solutions have smooth event horizons like multi-black holes 
with non-trivial asymptotic structure~\cite{Kimura:2008cq}.

To avoid conical singularities,  
the spacetime should be quantized by three relatively prime natural numbers $(p , p_+ , p_-)$  
satisfying the inequality $0 < p_+ - p_- < p < p_+ + p_-$.   
Then the topology of spatial infinity and that of each black hole 
are not S$^3$, but different lens spaces. 
The topology of spatial infinity 
is $L(p; q)$, 
while the topologies of the northern and the southern black holes are 
$L(p_+ ; q_+)$ and $L(p_- ; q_-)$, respectively, 
where 
$(p , q) ,~ (p_+ , q_+)$, and $(p_- , q_-)$ are three sets of coprime natural numbers.  
Then both the topologies and the geometries of the spatial infinity 
and black holes are uniquely determined 
by $(p , p_+ , p_-)$.

On the other hand,  
when the topology of spatial infinity 
is fixed as $L(p; q)$, 
there exist infinite sequences of coprime natural numbers $(p_+ , p_-)$ 
that give the same $q$. 
Using such sequences, 
we have infinite possibilities of lens space 
topologies of the northern and the southern black holes $L(p_\pm; q_\pm)$ for the fixed $(p , q)$.      
From equation \eqref{id_point}, we see that $(\sigma_1, \sigma_2)$, equivalently $(p_+, p_-)$, 
can take a variety of values 
if and only if both parameters $\alpha$ and $\nu$ do not vanish, otherwise, it should hold that $p_+=p_-$. 
Therefore, we can obtain the infinite number of possible lens space topologies of black holes, 
which are determined by $(p_+, p_-)$, in the case of Kerr-Taub-bolt base space.
The present spacetime is regular on and outside the black hole horizons, 
and represents a pair of Kaluza-Klein black holes with the rich variety of lens space 
topologies of horizons, 
even if the topology of spatial infinity is fixed.
We would explain the reason why the solution allows infinite possibilities of horizon topologies under a 
fixed topology of infinity in the framework of cobordism. However it is still an open question.

The solution \eqref{eq:metric} can be generalized to a solution 
of the system including a positive cosmological constant term $-4 \Lambda$ 
within the integral of the action \eqref{eq:action}
by replacing the harmonic function \eqref{hfunction2} into\footnote{
Introducing the new coordinate 
$t' = t - \sqrt{3 / 4 \Lambda}$
then taking the limit $\Lambda \to 0$, 
the metric \eqref{eq:metric} with the harmonic function \eqref{cosmoharmonic} 
represents the static solution \eqref{eq:metric} with the harmonic function \eqref{hfunction2}.  
}
\begin{eqnarray}
 H(t , r , \theta ) = \sqrt\frac{4 \Lambda}{3} t 
 + \frac{m _+}{r - \mu - (r_b - \mu) \cos \theta} 
 + \frac{m _-}{r - \mu + (r_b - \mu) \cos \theta}.
 \label{cosmoharmonic}
\end{eqnarray}
Similar to the cosmological solutions~\cite{Ishihara:2006ig, Matsuno:2007ts}, 
we expect that
the metric \eqref{eq:metric} with the harmonic function \eqref{cosmoharmonic} describes 
the physical process such that two black holes with the horizon topologies of $L(p_+ ; q_+)$ and $L(p_- ; q_-)$ 
coalesce into a single black hole with the horizon topology of $L(p ; q)$.
It would be interesting to compare the coalescence process between 
black holes on Kerr-Taub-bolt space 
and Gibbons-Hawking space~\cite{Ishihara:2006ig, Matsuno:2007ts}.\footnote{
The solutions \cite{Ishihara:2006ig, Matsuno:2007ts}
describe the coalescence of two black holes whose topologies are $L(p_1;1)$ and $L(p_2;1)$ 
into a single black hole with the topology of $L(p_1 + p_2;1)$~\cite{Yoo:2007mq, Kimura:2009gy}.}
We leave the analysis for the future.

\section*{Acknowledgments}
We would like to thank Benson Way and Yukinori Yasui for the useful discussions.  
This work is supported by the Grants-in-Aid for Scientific Research No. 19540305 and 
No. 24540282. 
MK is supported by a grant for research abroad from JSPS.

\appendix
\section{Lens space}
\label{appendix:lensspace}
\begin{figure}[htbp]
\centering
\includegraphics[width=7cm]{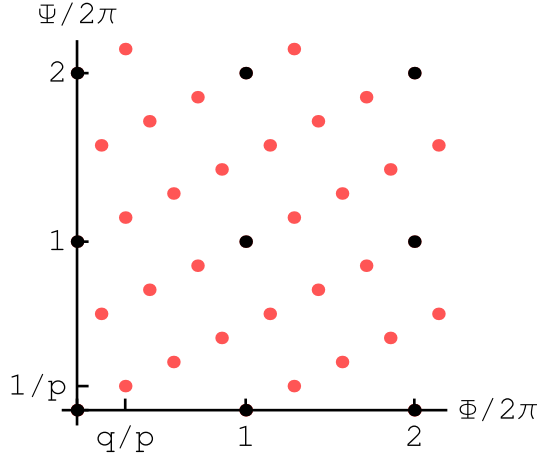}
\caption{
The identifications for $L(p;q)$.
As an example
we depicted the identical points same as $(\Phi, \Psi) = (0,0)$ for $L(7;2)$.
The black and red points come from the identifications (\ref{ids3}) and (\ref{idlens}), respectively.
The number of the points in the region $0<\Phi/2\pi<1$ and $0<\Psi/2\pi <1$ is $p-1$, 
and the number of possible lines with gradient $1/q$ through at least one of the points in this region is $q$.
}
\label{fig:lensspace}
\end{figure}

The metric of a unit ${\rm S}^3$ is given by
\begin{align}
ds^2 = d\Theta ^2 + \sin ^2 \Theta d\Phi ^2 + \cos ^2 \Theta d\Psi ^2,
\label{unitS3}
\end{align}
with the coordinate range $0 \le \Theta \le \pi/2$ and the identifications
\begin{align}
\left( \Phi, \Psi \right)\sim\left( \Phi + 2 \pi, \Psi \right) , 
\quad 
\left( \Phi, \Psi \right)\sim\left( \Phi, \Psi + 2 \pi \right).
\label{ids3}
\end{align}
A lens space $L(p ; q)$ is a manifold defined as the quotient of 
${\rm S}^3$ under the identification 
\begin{align}
\left(\Phi , \Psi \right) \sim \left( \Phi + 2 \pi \frac{q}{p}, \Psi + 2 \pi \frac{1}{p} \right),
\label{idlens}
\end{align}
where $p$ and $q$ are coprime natural numbers.\footnote{
The case of $p<0$ or $q < 0$ can be considered as a matter of the definition of signs of coordinates 
$\Phi$ or $\Psi$. Note that $q$ can also take $0$ only in the case of $p=1$.}
The lens space $L(p;q)$ becomes ${\rm S}^3$ in the case of $p = 1$, but
it has a different topology from ${\rm S}^3$ if $p \neq 1$.
It is known that two lens spaces $L(p; q)$ and $L(p^\prime; q^\prime)$ are diffeomorphic if and only if
the following two conditions are satisfied: 
(i) $p = p^\prime$, 
(ii) $(q \pm q' \bmod p) = 0$ or $(q q^\prime\pm 1 \bmod p ) = 0$.
For this reason, it is sufficient to consider the case $p > q$.
Fig.\ref{fig:lensspace} represents the identifications for $L(p;q)$.

Note that identical points same as the origin 
$(\Phi, \Psi) = (0,0)$ induced by the identifications (\ref{ids3}) and (\ref{idlens}) 
 are described as
\begin{align}
\left(\Phi , \Psi \right) =  \left( 2 \pi  \left(\frac{q}{p} a - b \right), 2 \pi \frac{a}{p} \right),
\label{appendix:kl}
\end{align}
where $a$ and $b$ are integers.
If $a$ and $b$ are coprime natural numbers, 
there is no identical point same as the origin
along the finite line connected with the origin and the point (\ref{appendix:kl}).
The area of a unit lens space $L(p; q)$ whose local metric is given by (\ref{unitS3}) becomes
\begin{align}
\mathcal A_{L(p;q)} = 
2 \pi^2/p.
\end{align}

If we use Euler coordinates $(\theta, \phi, \psi)$ defined by
\begin{align}
\Theta = \frac{\theta}{2} , \quad 
\Phi = \frac{\psi - \phi}{2} , \quad 
\Psi = \frac{\psi + \phi}{2},
\end{align}
the metric (\ref{unitS3}) becomes
\begin{align}
ds^2 = \frac{1}{4} \left[ d\theta ^2 + \sin ^2 \theta d\phi^2 
+ \left( d\psi + \cos \theta d\phi \right)^2 \right].
\label{unitS3euler}
\end{align}
In these coordinates, $0 \leq \theta \leq \pi$, and  
the identifications (\ref{ids3}) and (\ref{idlens}) become 
\begin{align}
&
\left(\phi, \psi \right)\sim\left(\phi \pm 2 \pi, \psi + 2\pi \right),
\label{ids3euler}
\\ &
\left( \phi , \psi \right) \sim \left(  \phi + 2 \pi(1-q) /p, \psi + 2 \pi(1+q)/p \right). 
\label{idlenseuler}
\end{align}

\section{Regularity condition on the bolt}
\label{appendix:boltregularity}

In this section, we study the regularity condition on the bolt off the black hole, i.e.,
$r = r_b$ and $\theta \neq 0, \pi$.
The metric of four-dimensional Kerr-Taub-bolt space \eqref{base} diverges at $r = r_b$ apparently.  
These points correspond to fixed points of the Killing vector field 
$\partial/\partial \psi
+ 
2 \alpha \nu/(r_b^2 - \nu ^2 - \alpha ^2) 
\partial/\partial \phi$.
We expand the metric \eqref{eq:metric} near these fixed points, 
and obtain regularity conditions of identifications.

Introducing new coordinates,  
\begin{align}
 R = \sqrt \frac{2 (r - r_b)}{r_b - \mu} ,
\quad
 \chi = \frac{- 2 \alpha \nu}{r_b^2 -\nu ^2 -\alpha ^2} \psi + \phi ,
\end{align}
near $r = r_b$, the metric \eqref{eq:metric} behaves as  
\begin{align} 
ds^2 &\simeq&  - H(r_b,\theta )^{-2}dt^2
+ H(r_b,\theta ) \Xi(r_b,\theta ) 
\bigg[
dR^2 + R^2 
\bigg( \frac{r_b - \mu}{\Xi (r_b,\theta ) }(2 \nu \cos \theta + \alpha \sin ^2 \theta) d\chi 
\notag \\
& &
+ \frac{2 \nu (r_b - \mu)}{r_b^2 - \nu^2 - \alpha ^2} d\psi \bigg) ^2 
+ 
 d\theta^2 + \frac{(r_b^2 - \nu^2 - \alpha ^2) ^2}{\Xi (r_b,\theta ) ^2} \sin ^2 \theta d\chi^2 
\bigg] .
\label{nearboltbase}
\end{align}
The induced metric on the two-dimensional bolt  
is given by the second and the third terms 
in the second line of the above metric.
The function $\Xi(r_b, \theta) = r_b^2 - (\nu - \alpha \cos\theta)^2$ takes 
larger value in the north part than the south part,
so the shape of the bolt is like a pear. 
In fact, this asymmetric behavior leads a difference of topologies of each black holes.

Near the bolt $R = 0$, 
the metric \eqref{nearboltbase} with $t, \theta, \chi = {\rm const.}$ takes the form  
\begin{align} 
ds^2|_{t,\theta,\chi = {\rm const.}} \simeq  
H(r_b,\theta )
\Xi (r_b,\theta ) 
\left[   
dR^2 + R^2 \left(\frac{2 \nu (r_b -\mu)}{r_b ^2 -\nu ^2 -\alpha ^2} d \psi \right)^2 
\right] .
\end{align}
The above two-dimensional metric is regular if and only if 
the coordinate $\psi$ is periodic with a period $2 \pi (r_b ^2 -\nu ^2 -\alpha ^2)/(2 \nu (r_b -\mu))$
along $\chi = - 2 \alpha \nu \psi/(r_b^2 -\nu ^2 -\alpha ^2) + \phi =
{\rm const.}$ in $(\phi, \psi)$ plane.
Equivalently, such the condition is same as the following identification
\begin{align}
\left( \phi, \psi \right) \sim 
\left( \phi + 2 \pi \frac{\alpha  }{r_b - \mu}
, \psi + \pi \frac{r_b^2 - \nu^2 - \alpha^2}{\nu(r_b - \mu)} \right).
\label{npbhreg2appendix}
\end{align}
Using the coordinates \eqref{global_coord}, 
we see that the regularity condition \eqref{npbhreg2appendix} coincides with 
the condition \eqref{SP_reg}.


\end{document}